\documentclass[a4paper,11pt]{article}
\usepackage{jcappub} 
\usepackage{lineno}
\usepackage{siunitx}
\usepackage{rotating}
\usepackage{multirow}
\usepackage{makecell}

\title{\boldmath Dependence of the estimated electric potential in thunderstorms observed at GRAPES-3 on the hadronic interaction generators used in simulations}

\author[a,1]{B. Hariharan\note{Corresponding author}}
\author[a]{S.K. Gupta}
\author[b]{Y. Hayashi}
\author[a]{P. Jagadeesan}
\author[a]{A. Jain}
\author[b]{S. Kawakami}
\author[c]{H. Kojima}
\author[a]{P.K. Mohanty}
\author[d]{Y. Muraki}
\author[a]{P.K. Nayak}
\author[c]{A. Oshima}
\author[a]{M. Rameez}
\author[a]{K. Ramesh}
\author[a]{L.V. Reddy}
\author[c]{S. Shibata}
\author[a]{M. Zuberi}

\affiliation[a]{Tata Institute of Fundamental Research, Homi Bhabha Road, Mumbai, 400005, India}
\affiliation[b]{Graduate School of Science, Osaka City University, Osaka, 558-8585, Japan}
\affiliation[c]{College of Engineering, Chubu University, Kasugai, 487-8501, Japan}
\affiliation[d]{Institute for Space-Earth Environmental Research, Nagoya University, Nagoya, 464-8601, Japan}

\emailAdd{89hariharan@gmail.com}

\abstract{A potential difference of 1.3 Giga-Volts (GV) was inferred across a thundercloud using data from the GRAPES-3 muon telescope (G3MT) \textbf{\cite{Hari_2019_1}}. This was the first-ever estimation of gigavolt potential in thunderstorms, confirming prediction of C.T.R. Wilson almost a century ago. To infer the thundercloud potential required acceleration of muons in atmospheric electric field to be incorporated in the Monte Carlo simulation software CORSIKA. The G3MT records over 4 billion muons daily that are grouped into 169 directions covering 2.3 sr sky. This enabled changes as small as 0.1\% in the muon intensity on minute timescale, caused by thunderstorms to be accurately measured. But that requires high statistics simulation of muons in thunderstorm electric fields. The CORSIKA offers a choice of several generators for low- (FLUKA, GHEISHA, and UrQMD) and high-energy (SIBYLL, EPOS-LHC, and QGSJETII) hadronic interactions. Since it is unclear which combination of the low- and high-energy generators provides the correct description of hadronic interactions, all nine combinations of generators were explored, and they yielded thundercloud potentials ranging from 1.3\,GV to 1.6\,GV for the event recorded on 1 December 2014. The result of SIBYLL-FLUKA combination yielded the lowest thundercloud potential of 1.3\,GV was reported. Furthermore, another seven major thunderstorm events recorded between April 2011 and December 2020 were analyzed to measure the dependence of their thundercloud potential on the hadronic interaction generators. It is observed that the low-energy generators produce larger variation ($\sim$14\%) in thundercloud potential than the high-energy generators ($\sim$8\%). This probably reflects the fact that the GeV muons are predominantly produced in low-energy ($<$80\,GeV) interactions, which effectively magnifies the differences in the meson production cross-sections among the low-energy generators.}

\newcommand{\imu}{$\Delta$I$_{\mu}$}
\date{}
\begin{document}
\maketitle
\flushbottom

\section{Introduction} 

Thunder and lightning, although known to humanity for millennia, are feared for their destructive consequences. The first documented studies were conducted independently in Europe and North America by Thomas-Fran\c cois Dalibard and Benjamin Franklin, respectively, in the 1750s. These studies demonstrated the electrical nature of thunderstorms \cite{Web_1750,Franklin_1752}. Armed with this understanding, Franklin went on to invent and popularize the lightning arrester to protect buildings. The electrification of thunderclouds is understood as follows: moist hot air moves upward in an updraft. Then the temperature gradient in the atmosphere leads to rapid cooling of the air, producing water droplets, ice crystals, and hail pellets (graupel). The updraft continues to carry the water droplets and ice crystals. However, the progressively heavier graupel can not be moved by the updraft, and first it stays suspended in the atmosphere, but beyond a certain mass starts to fall. The collisions among upward-moving water droplets and falling graupel leads to charge exchange. The positively charged water droplets/ice crystals reach the top of the cloud, whereas the negatively charged graupel resides at the bottom of the cloud. This process continues as long as the electrical insulation in the cloud can withstand the electric potential generated by charge separation. The eventual breakdown of air insulation leads to discharge that produces bursts of lightning. This electrification of thunderclouds is dependent on a variety of factors such as ambient temperature, humidity, wind speed, the size of ice crystals, etc. Typical thunderclouds are known to have charged layers separated by several kilometers (km). The thickness of the charged layers is typically a few kms. Of course, in reality, thunderclouds are far from simple dipoles; they are known to exhibit intricate structures with multiple charged layers \cite{Mason_1972,Mason_1988,Mason_2003,Williams_1988,Saunders_2008}.

C.T.R. Wilson had measured electric field strengths of up to 5$\times$10$^6$\,V/m and knowing that the charged layers of thunderclouds extend over several kilometers, he had estimated that potential differences of about a gigavolt could be generated \cite{Wilson_1924,Wilson_1929,Wilson_1956}. Subsequently, there have been sustained attempts to measure potential difference across the thunderclouds by using balloon and rocket soundings. The weather balloons can carry payloads of electric field meters up to altitudes of 30--40 km. These balloons are typically launched during turbulent weather conditions ripe for thunderstorm formation, and electric field measurements are made at various heights in the atmosphere. The measured electric field is then integrated to estimate the thundercloud potential. Similar measurements have also been made by using sounding rockets. The sounding rockets typically fly for about a minute, whereas weather balloons allow recordings are made from a few tens of minutes to hours. To date, the highest estimates of thundercloud potential from the sounding technique is 0.13\,GV \cite{Marshall_1995,Marshall_2001}. There have been reports of the detection of MeV gamma-rays on the ground during lightning from thunderstorms \cite{Dwyer_2003,Ringuette_2013}. Thunderstorms have been identified as the source of terrestrial gamma-ray flashes (TGFs), discovered by the Burst And Transient Source Experiment (BATSE) aboard the ``Compton Gamma Ray Observatory'' \cite{Fishman_1994}. Such gamma-rays are produced through the bremsstrahlung of electrons of energies about an order of magnitude higher, which suggests the presence of large potentials in the TGFs. Therefore, a thundercloud potential of 0.13\,GV would be inadequate to produce gamma-rays of 40--50\,MeV detected in the TGFs, not to mention the detection of 100\,MeV gamma-rays reported by the instruments aboard Astro-rivelatore Gamma a Immagini Leggero (AGILE) satellite \cite{Tavani_2011}.

In 2019, the \textbf{G}amma \textbf{R}ay \textbf{A}stronomy at \textbf{P}eV \textbf{E}nergie\textbf{S} -- phase \textbf{3} (\textbf{GRAPES-3}) reported the estimation of a massive thundercloud potential of 1.3\,GV \cite{Hari_2019_1} in an event recorded on 1 December 2014. This potential was obtained through the variation in the measured muon intensity (E$_\mu \geqslant$1\,GeV). The earlier studies of low-energy muons (90\,MeV) by several groups had indicated a close connection between the atmospheric electric field and variations in muon intensity \cite{Alexeenko_1987,Dorman_2003}, which were later confirmed by the experiment on Mount Norikura \cite{Muraki_2004}. These earlier observations had revealed that atmospheric muons can be an effective tool for studying the electrical properties of thunderclouds. Unlike the sounding technique that can monitor the electric fields in thunderclouds in a relatively small area and for only a brief duration of a few hours, the GRAPES-3 muon telescope (G3MT) monitors a large area of the sky (hundreds of km$^2$) round the clock. The muons detected by the G3MT are produced in extensive air showers (EAS), arising from the interactions of cosmic rays (CRs) in the atmosphere, and a vast majority ($\sim$90\%) of them are produced well above the typical height of cumulonimbus clouds ($\sim$10\,km above mean sea level ``amsl''). Since muons are electrically charged, they are influenced by the electric field within the thunderclouds. Due to their relativistic speeds, muons traverse the thunderclouds in just a few microseconds, and consequently provide information on their electrical properties on time-scales shorter than the lifetime of thunderclouds.

However, the interpretation of variations in the observed muon intensity requires detailed Monte Carlo simulations of the EAS development and the detector response. The thunderstorm events observed by the G3MT were studied by relying on CORSIKA simulations to model the muon production in the atmosphere after incorporating a thundercloud model to reproduce the observed muon intensity variations. These simulations were performed as the function of a wide range of thundercloud potentials and several hadronic interaction generators implemented in CORSIKA. The thundercloud structure in CORSIKA is implemented by defining two charged layers of thickness of 2\,km each, which are also separated by 2\,km. The choice of thundercloud structure is not critical because the variation of muon intensity primarily depends on the potential across the thundercloud and is not mostly insensitive to its structure. Thus, the scope of this report is limited to probing the effects of different hadronic interaction generators on the estimate of thundercloud potential. With the aid of these simulations, the thunderstorm event observed on 1 December 2014 was analyzed in greater detail and the properties of this event were derived featuring the first-ever estimation of a thundercloud potential of 1.3\,GV as reported in \cite{Hari_2019_1}. 

The atmospheric electric field (V/m) is a crucial physical parameter in the study of thunderstorms and associated effects. An electric field mill (EFM) is a simple yet sensitive device that operates on the principles of a capacitance electrometer. EFMs are widely used for monitoring local and ambient atmospheric electric fields through the charge accumulated on a surface area \cite{Antunes_2020}. We have installed four Boltek EFM-100 Atmospheric Electric Field Monitors to observe the atmospheric electric field around GRAPES-3 \cite{Boltek}. Figure\,1c of \cite{Hari_2019_1} shows the deployment of these four EFMs. These instruments have been in continuous operation since April 2011, collecting data with a resolution of 0.01\,V/m every 50\,milliseconds.  The measured atmospheric electric field from these four widely spaced EFMs and combined with muon measurements by G3MT, the movement of the thundercloud could be tracked including its altitude, speed, and area. During the course of these investigations it became clear that the estimated thundercloud potential was sensitive to the choice of hadronic interaction generators used in the CORSIKA simulations, and its value varied between 1.3\,GV and 1.6\,GV. A conservative approach was adopted by selecting the SIBYLL-FLUKA combination, which yielded the lowest estimated peak potential of 1.3\,GV. Furthermore, in the current work, we have extended this analysis to include seven more major thunderstorm events recorded between April 2011 and December 2020 to assess the effect of hadronic interaction generators in a more general manner because of its significance in future studies of thundercloud properties. Results of these investigations of hadronic interaction generators are summarized in this work. The details of G3MT and muon direction reconstruction are discussed in Section\,\ref{G3}. Section\,\ref{MC} provides the details of Monte Carlo simulations, followed by a brief description of the analysis of thundercloud events recorded by G3MT in Section\,\ref{Event}. A detailed account of the dependence of the estimated thundercloud potential on different combinations of low- and high-energy hadronic interaction generators is provided in Section\,\ref{Model}. Finally, the observations are discussed in Section\,\ref{Discussion}, and the manuscript is concluded with a brief summary in Section\,\ref{Summary}.


\section{The GRAPES-3 muon telescope (G3MT) \label{G3}} 

The GRAPES-3 experiment is designed for the study of EASs produced by CRs around the `knee' in their energy spectrum. GRAPES-3 is located at Ooty in southern India (11.4$^\circ$N, 76.7$^\circ$E, 2200\,m amsl). Its near-equatorial location provides excellent sky coverage of the northern and southern hemispheres of the sky. The GRAPES-3 experiment consists of two distinct detector components: (i) a high-density EAS array of nearly 400 plastic scintillators (each 1\,m$^2$ area) detectors (G3SD) spread over 25000\,m$^2$ and (ii) a large area (560\,m$^2$) muon telescope (G3MT). The details of the detectors, the logic used in trigger generation, the data acquisition system, etc. can be found at \cite{Gupta_2005}. The G3SD records more than 3.5$\times$10$^6$ EAS every day. The second detector component, the G3MT is used to measure the muon content of the EAS. The basic building blocks of G3MT are durable gas-filled proportional counters (PRCs). Each PRC is made from a 600\,cm long hollow steel tube of wall thickness 2.3\,mm with a square cross-section of 10$\times$10\,cm$^2$. The PRCs are sealed at both ends and filled with a P10 gas mixture (90\% argon and 10\% methane) at a pressure of approximately 25\% above the local atmospheric pressure. At the center of the tube, a 100\,\si{\micro\meter} diameter tungsten wire serves as anode, electrically isolated from the steel body by using hermetic seals at both ends of the PRC. A potential difference of about +3000\,V$_\text{DC}$ is applied between the anode and the cathode, providing sufficient gain for the detection of a singly charged relativistic particle, such as a muon.

When a charged particle passes through the PRC, electron-ion pairs are produced by the ionization of gas molecules. The electrons are multiplied by the presence of a large electric field around the anode that causes local avalanches, which are then quenched by the methane present in the P10. The height of the pulse received at the anode is proportional to the energy deposited by the passing particle. The PRCs that make up G3MT are divided into sixteen independent modules. Each module comprises four layers with 58 PRCs in each layer that are tightly packed next to each other and the alternative layers are arranged orthogonal to each other. The adjacent layers are separated by a 15\,cm thick layer of concrete blocks. Above the uppermost layer, 2\,m thick concrete blocks are placed in the shape of an inverted pyramid to provide a coverage of up to 45$^\circ$ for inclined muons. The total mass overburden of the concrete of $\sim$550\,g/cm$^2$ yields an energy threshold of sec($\theta$)\,GeV for the muons of incident angle $\theta$. The sixteen muon modules cover a total area of 560\,m$^2$ making it the world's largest muon telescope with a mean angular resolution of about 4$^{\circ}$. The G3MT is equipped with two distinct data acquisition (DAQ) systems. The first records the PRC hits that are associated with the EAS triggers generated by G3SD. The offline reconstruction of muon directions from this data serves as the primary tool to discriminate the EAS produced by gamma-ray primaries from hadronic primaries, as well as to measure the nuclear composition of the primary CRs.

\begin{figure*}[t]
    \centering
    \includegraphics*[width=0.95\textwidth]{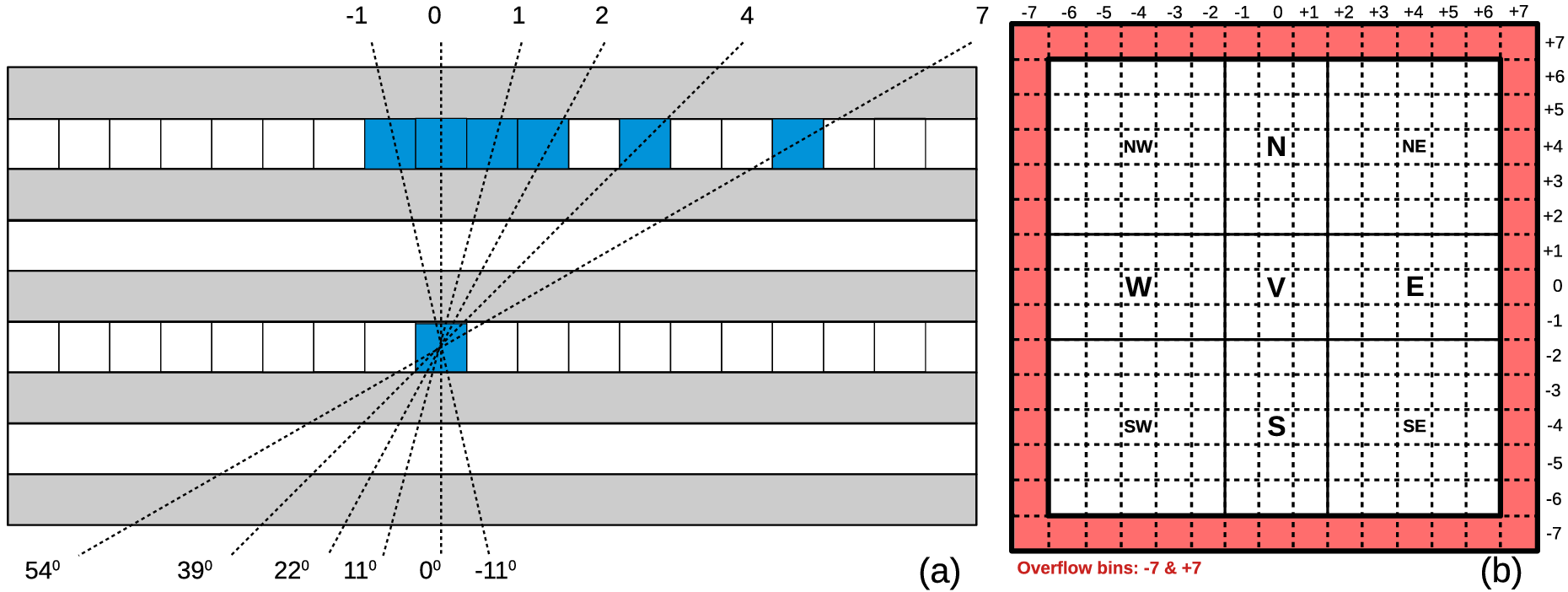}
    \caption{Representation of (a) muon angle reconstruction in a single projection using PRC hits and (b) G3MT's sky coverage in 169-direction configuration (excluding outermost overflow bins) shown as dotted lines. The solid lines represent coarser 9-direction configuration.}
    \label{Fig_1}
\end{figure*}

The second DAQ is used to record the directional muon intensity in the absence of an EAS trigger. This DAQ relies on a logic system that generates a trigger whenever a coincidence of PRC hits is recorded in all four layers (4F) of a muon module. Next, the PRC hits associated with 4F are latched, and the arrival time and pulse width information of all hits are transferred to the DAQ PC. The orthogonal placement of alternate layers of PRCs allows the reconstruction of muon direction in two projection planes: XZ and YZ. In each projection, the spatial separation of the hit PRCs in the upper and lower layers is calculated as shown in Figure\,\ref{Fig_1}a (i.e. vertical muon defined as direction 0). The inclined muons are selected up to a separation of --7 to +7 PRCs, for a total of 15 directions in each projection plane. The muons inclined beyond the 7$^{th}$ PRC are also stored in the direction of 7$^{th}$ PRC, consequently becomes the overflow direction. By combining the information from the two projections, the muon direction in the sky is binned into one of 225 directions (15$\times$15) covering 2.3\,sr as shown in Figure\,\ref{Fig_1}b. However, for further data analysis, the overflow directions are not used, thereby only information from the inner grid of 13$\times$13\,=\,169 directions are used. The DAQ software integrates the reconstructed muon directions into 10-second live-time bins before storing the data. Depending on the nature of the physics problem, all 169 directions can be used or further combined into coarser 9 directions as shown in Figure\,\ref{Fig_1}b. The muon intensity is first corrected for variations in atmospheric parameters and the efficiency of the detector \cite{Mohanty_2016_2,Mohanty_2017_2}. The G3MT records about four billion muons per day that are predominantly produced by CRs of energy from 10\,GeV to 10\,TeV. The studies of angular muon intensity have proven to be an excellent probe for atmospheric and near-Earth phenomena \cite{Hari_2019_1,Mohanty_2016_1,Mohanty_2017_1,Hariharan_2023,Hariharan_2022}. The details of the G3MT can be found in \cite{Hayashi_2005}.


\section{Monte Carlo simulations \label{MC}}

Interpreting ground-based CR data requires detailed Monte Carlo simulations of EAS development in the atmosphere. CORSIKA (\textbf{CO}smic \textbf{R}ay \textbf{SI}mulations for \textbf{KA}scade) is a widely used simulation software by the global scientific community for the studies of EAS development. It was originally developed and is maintained by the CR group at Karlsruhe Institute of Technology, Germany. The entire CORSIKA package contains more than 80000 lines of \texttt{FORTRAN} code with some optional subroutines in \texttt{C}. It enables the study of the EAS development of primary gamma-rays and nuclei of different elements in the atmosphere. The secondary particles produced in the interaction of primary CRs in the atmosphere are tracked up to a user-defined observational level and energy thresholds for different particles. At the observational level, the output file records the position (X, Y, Z), momentum (P$_X$, P$_Y$, P$_Z$), and the time of flight (t) from the first interaction point. The electromagnetic cascades are handled by EGS4 \cite{egs4} code or by analytical Nishimura-Kamata-Greissen (NKG) \cite{nkg} functions according to the user's choice.  The hadronic interactions in the high-energy domain can be handled by one among several external generators that have been incorporated into CORSIKA such as EPOS-LHC \cite{eposlhc}, QGSJET01C \cite{qgsjet}, QGSJETII-04 \cite{qgsII}, SIBYLL\,2.1 \cite{sibyll_2.1}, etc.  Similarly, low-energy interactions are managed by using one of the following generators: FLUKA\,2011-2B \cite{fluka}, GHEISHA\,2002d \cite{gheisha}, or UrQMD\,1.3cr \cite{urqmd}. A combination of low- and high-energy hadronic interaction generators, chosen by the user, is used for the simulations. Further technical details of CORSIKA can be found in \cite{corsika}.

During the simulations, the atmospheric conditions are mimicked by selecting from a set of pre-defined atmosphere models. These models divide the atmosphere into five layers. The first four atmospheric layers are modeled with an exponential dependence on altitude. The uppermost layer is modeled using a linear function because of the extremely low density of air at that altitude. Here, an in-built atmospheric model of ATMOD-5, corresponding to AT616 Central European atmosphere for June 16, 1993 is used. To simulate the effects of thunderstorms in CORSIKA, an in-built option called `EFIELD' may be used. However, this option was originally implemented only for the electromagnetic component. Additionally, in the EFIELD option, the applied electric field is assumed to be uniform for the entire depth of the atmosphere. However, in reality, thunderclouds possess a limited thickness in the range of a few kms. Therefore, we modified the CORSIKA code by extending the scope of EFIELD to include the pions, kaons, and muons \cite{Hari_2017_2}, and the electric field is applied over a limited pre-defined depth of the atmosphere to mimic charged layers of a thundercloud. The authors then added the EFIELD functionality to other particles in the more recent versions of CORSIKA due to the increased interest in this application within the CR community. In a nutshell, the change in the energy of a charged particle is calculated for the path length traversed during its transport through the thundercloud, which is then either added or subtracted from the initial energy of the particle depending on the polarity of its charge. 

\begin{figure*}[t]
    \centering
    \includegraphics*[width=0.90\textwidth]{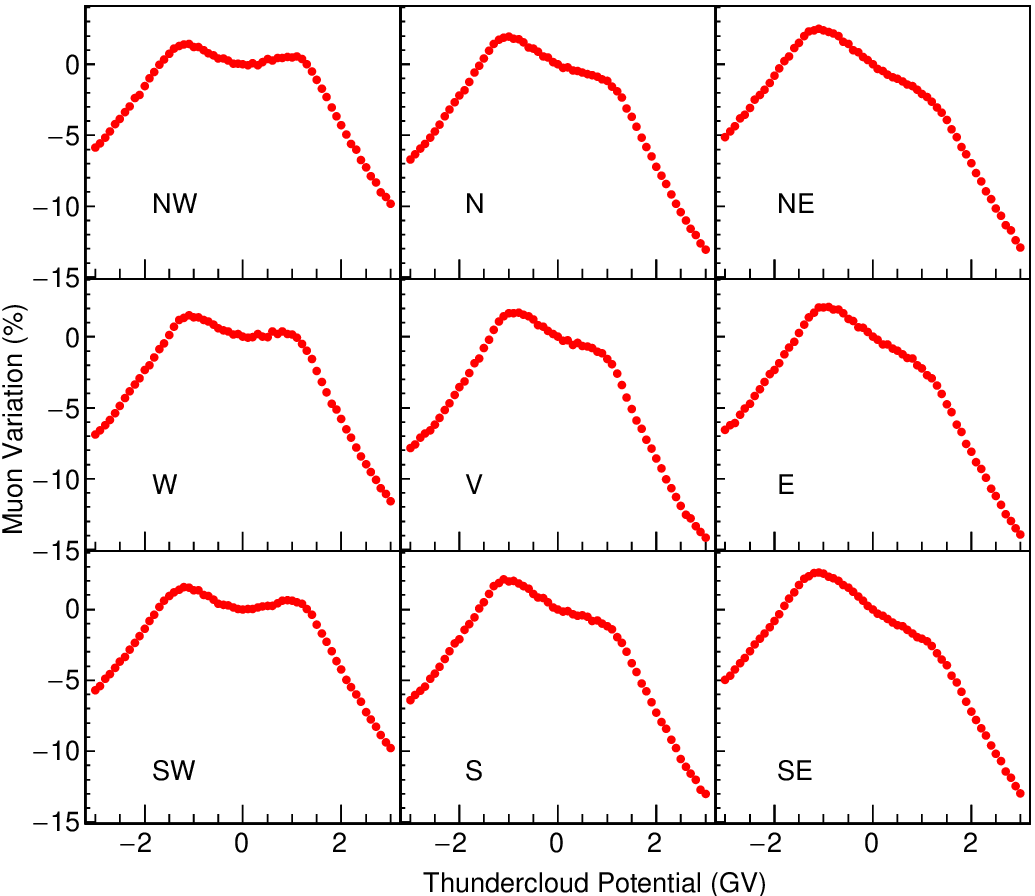}
    \caption{Simulated profiles of percent variation in muon intensity (\imu) as a function of the applied thundercloud potential, shown for 9-direction configuration. In the eastern (NE, E, SE) and the central directions (N, V, S), the \imu{} decreases for \textbf{+V} up to $\sim$1\,GV due to the well-known fact of muon charge asymmetry ($\mu_R=\frac{N_{\mu^+}}{N_{\mu^-}}>$1) and vice versa, whereas the western directions (SW, W, NW) display weaker dependence due to value of $\mu_R$ approaching 1. The \imu{} decreases rapidly beyond an applied potential of 1\,GV due to the enhanced probability of the decay of GeV muons irrespective of the polarity of applied potential.}
    \label{Fig_2}
\end{figure*}

CORSIKA-v74001 has been used to study the effect of thunderstorms on the muon intensity \cite{Hari_2019_1}. For these simulations, an energy range of 10\,GeV--10\,TeV for primary CRs was selected. This is because the CRs in this energy range are responsible for producing an overwhelming fraction of muons detected by the G3MT. Since the primary composition in this energy range is predominantly composed of protons ($\sim$90\%), the simulations were restricted to protons with a spectral index of --2.65. This index is obtained from a combined fit to the proton energy spectra measured by PAMELA, CAPRICE, BESS, and CREAM \cite{Chandra_2015}. The efficiency of the selection of proton primaries was enhanced by an innovation which we introduced in CORSIKA, namely, the CRs were simulated only if the rigidity of the primary CR was higher than 90\% of the cutoff rigidity corresponding to that direction. This constraint resulted in a significant reduction in the computation time required for simulations, especially because G3MT has a threshold that is comparable to the cutoff rigidity. A reduction in computing time by a factor of $\sim$3 was obtained \cite{Hari_2015,Hari_2019_3}.

The muon intensity in the field of view (FOV) of G3MT reduces rapidly from the vertical direction to more inclined directions due to its steep dependence on zenith angle. To compensate for the loss of statistics due to this zenith angle dependence, the primaries were selected in the center of each direction bin instead of following the normal procedure of simulating the CRs isotopically. For estimating the background, the number of primaries in each of the 169 directions were appropriately scaled to result in a detection of 10$^7$ muons, resulting in a statistical error of only 0.03\%. A wide range of thundercloud potentials, varying from --3\,GV to +3\,GV in steps of 0.1\,GV were simulated with CORSIKA, after modification to incorporate electric potential in a thundercloud. Each value of potential was implemented by applying a uniform electric field of required magnitude and direction between the altitudes of 8--10\,km amsl, simulating 10$^6$ muons in each direction (statistical error of 0.1\%). This particular choice of thundercloud altitude is reasonable and is consistent with the observations made by the G3MT \cite{Hari_2019_1}. Since these simulations require large computational resources, this strategy was devised to create a ready-made bank of simulated muons with high statistics that could be efficiently used to measure the effects of a thundercloud. This data bank was generated by selecting SIBYLL and FLUKA for the high- and low-energy hadronic interaction generators, respectively.

The percent variation in muon intensity (\imu) as a function of the applied thundercloud potential is derived for all 169 directions. But only the variation of \imu{} for the 9-direction configuration is shown in Figure\,\ref{Fig_2} for ease of understanding. The simulated profiles display several distinct features. In the three east (NE, E, SE) and the three central directions (N, V, S), the \imu{} decreases with \textbf{+V} up to $\sim$1\,GV due to the well-known fact of muon charge asymmetry ($\mu_R=\frac{N_{\mu^+}}{N_{\mu^-}}>$1). However, this effect is dominant in the six directions mentioned above due to a larger value of $\mu_R$, particularly when compared to the three western directions (NW, W, SW) as seen in Figure\,\ref{Fig_2}. The muons detected by the G3MT are predominantly produced by the CRs with energies in the range of few tens of GeV. As a result, \imu{} decreases rapidly beyond an applied potential of 1\,GV due to the increased probability of the decay of GeV muons irrespective of the polarity of the applied potential \cite{Hari_2019_1,Hari_2017_2,Alexeenko_1985,Alexeenko_1987}. It is also important to note that the overall decay rate is significantly higher for \textbf{+V} than for \textbf{--V} as may be seen in Figure\,\ref{Fig_2}. The change in \imu{} observed in the simulated profiles can be used to estimate the thundercloud potential in the respective directions.


\section{Thunderstorm events recorded by G3MT \label{Event}}

\begin{figure*}[t]
    \centering
    \includegraphics*[width=0.90\textwidth]{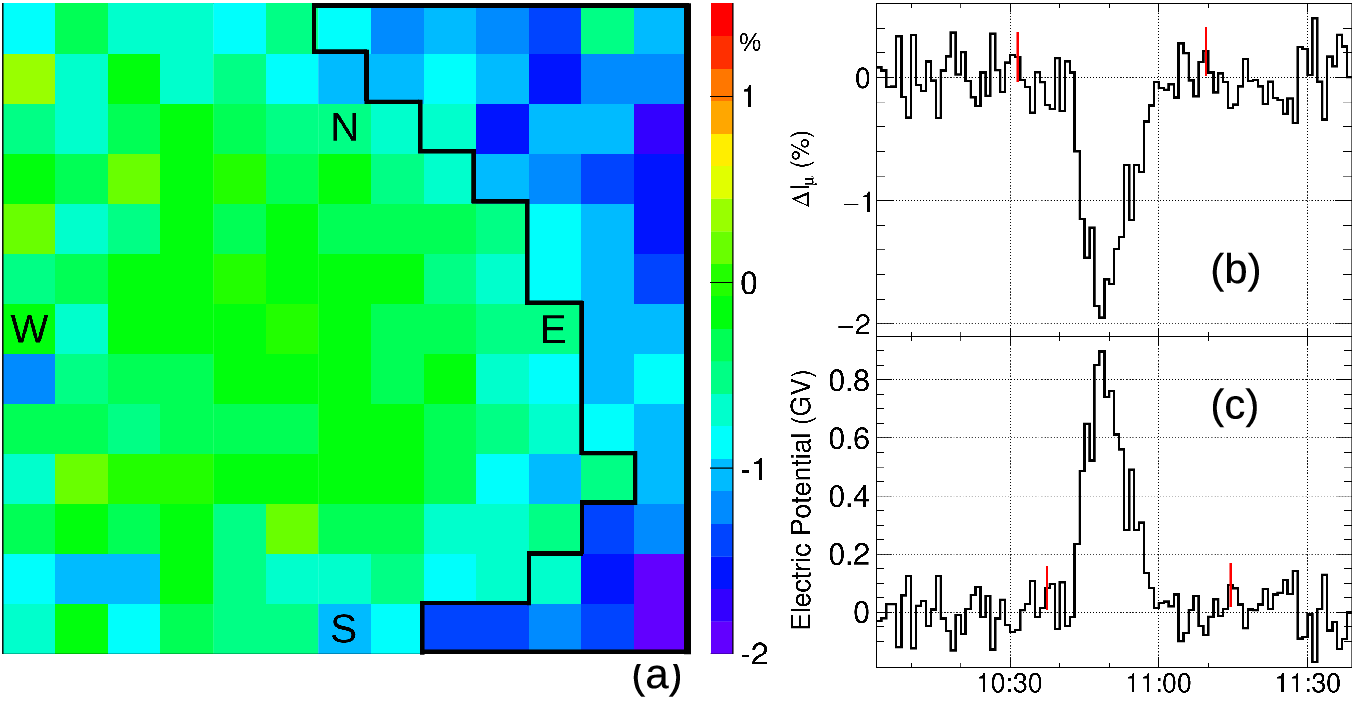}
    \caption{(a) Muon image of 1 December 2014 thunderstorm event lasted 18 minutes. In total 45 out of 169 contiguous directions enclosed by a dark boundary show significant variation. The percent variation is displayed in the adjacent colour coded bar; (b) The time variation of muon intensity (\imu{}) derived from 45 contiguous directions shows a peak deficit of (2.0$\pm$0.2)\%; (c) The profile shows thundercloud potential derived from Monte Carlo simulations. The peak potential is found to be (0.90$\pm$0.08)\,GV at 10:48 UT. The vertical bars inside the figures represent $\pm$1$\sigma$ error. Figures are reproduced from \cite{Hari_2019_1}.}
    \label{Fig_3}
\end{figure*}

In a recent work, several important characteristics of a major thunderstorm recorded by the GRAPES-3 experiment on 1 December 2014 that lasted about 18 minutes were reported \cite{Hari_2019_1}. The muon intensity recorded during this event showed a significant deficit in 45 contiguous directions out of a total of 169 directions. This deficit spanned most of the eastern region with a small presence extending into the northern and southern regions as shown in Figure\,\ref{Fig_3}a. The muons in these contiguous directions were combined and the resulting peak deficit of (2.0$\pm$0.2)\% corresponds to a significance of 10$\sigma$, however, if the effect over the entire 18-minute period is considered, it has a combined significance of about 20$\sigma$ as shown in Figure\,\ref{Fig_3}b. This deficit in muon intensity can be used to estimate the potential across the thundercloud. This is done by combining the simulated profiles of selected 45 directions to reproduce the observed deficit. The peak deficit of 2\% corresponds to a peak thundercloud potential of (0.90$\pm$0.08)\,GV as seen in Figure\,\ref{Fig_3}c. Further studies with muon imaging and a shorter 2-minute exposure window demonstrated clear evidence of the thundercloud's movement and resulted in an estimate of 1.3\,GV for the thundercloud potential. This was the first-ever estimation of gigavolt potential in a thundercloud. The possibility of thunderclouds generating gigavolt potential was first predicted by C.T.R. Wilson nearly a century ago \cite{Wilson_1924,Wilson_1929,Wilson_1956}.

The GeV muons detected by the G3MT, being relativistic particles, and combined with the fact that they almost traverse through the entire depth of the atmosphere, make them ideal candidates for thunderstorm studies. This may also explain the lack of success of sounding experiments to measure thundercloud potentials beyond 0.13\,GV due to their limited coverage area and live-time \cite{Marshall_2001}. In contrast, G3MT operates throughout the year with a sky coverage of 2.3 sr, which monitors a huge volume of the atmosphere with nearly 100\% duty cycle. Although the G3MT has been operational for more than two decades, thunderstorm events could be studied in great detail after the installation of EFMs in April 2011. A collection of 487 significant thunderstorm events recorded from 2011 to December 2020 comprises a high-statistics dataset of observations with an annual average of about 50 thunderstorm events \cite{Hariharan_2021}. \textbf{In this study, we analyze seven additional major thunderstorm events identified from the extended dataset corresponds to the above-mentioned period and estimate the corresponding thundercloud potentials. Several events exhibit gigavolt-scale potentials, indicating that such high-magnitude potentials are a characteristic feature of these thunderstorms. Since these estimates are derived through Monte Carlo simulations, it is crucial to quantitatively assess the differences among various hadronic event generators. To achieve this, a statistical analysis of the eight thunderstorm events is conducted to evaluate the overall variations introduced by different hadronic interaction models.}


\section{Dependence on hadronic interaction generators \label{Model}}

The estimation of gigavolt thundercloud potential is the result of a precise estimate of changes in muon intensity (\imu{}) and its subsequent interpretation with the aid of detailed Monte Carlo simulations. These simulations suggest that the estimated potential is dependent on the interaction generators employed. The results discussed in Section\,\ref{MC} are based on SIBYLL and FLUKA for high- and low-energy hadronic interaction generators, respectively. The CORSIKA is integrated with a handful of popular hadronic interaction generators. Although these generators are based on a simple parton model of Gribov-Regge multiple scattering, the results differ due to different approaches in the implementation of physics treatment, as well as the approximations used in describing the hadronic interactions \cite{Pierog_2017}. The values of various interaction cross-sections and particle multiplicities are based on the available data from the collider experiments, which are extrapolated to higher energies of the CRs. It is the implementation of physics among these generators via different approaches that leads to variations in the production of pions, kaons, and muons, which is a key cause of the observed difference among the predictions of these generators. Hence, the electrical properties in turn derived from the simulations display variations depending on the choice of generators. To study this dependence, nine combinations of the widely used generators, namely, SIBYLL, EPOS-LHC, QGSJETII (E$\,>\,$80\,GeV) and FLUKA, GHEISHA, UrQMD (E$\,<\,$80\,GeV) are selected. The SIBYLL-FLUKA combination was already available with large statistics, as discussed in Section\,\ref{MC}. Further simulations were carried out for the remaining combinations by following the same strategy but with reduced statistics. The primaries were scaled appropriately to produce 10$^6$ muons for background and 10$^5$ muons for signal in each of the 169 directions. The derivation of \imu{} from the simulation remains the same as discussed in Section\,\ref{MC}. This scheme allowed us to explore the dependencies brought about by the choice of hadronic interaction generators. 

\begin{figure*}[t!]
    \centering
    \includegraphics*[width=0.95\textwidth]{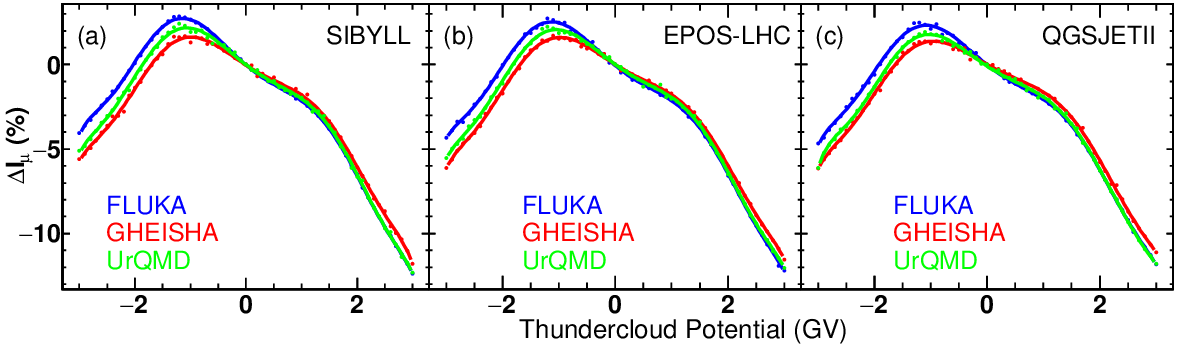}
    \caption{Simulated profiles of percent variation in muon intensity (\imu{}) among low-energy hadronic interaction generators i.e. FLUKA, GHEISHA, and UrQMD are shown with reference to common high-energy hadronic interaction generator (a) SIBYLL, (b) EPOS-LHC, and (c) QGSJETII.}
    \label{Fig_4}
    \vspace{1cm}
    \includegraphics*[width=0.95\textwidth]{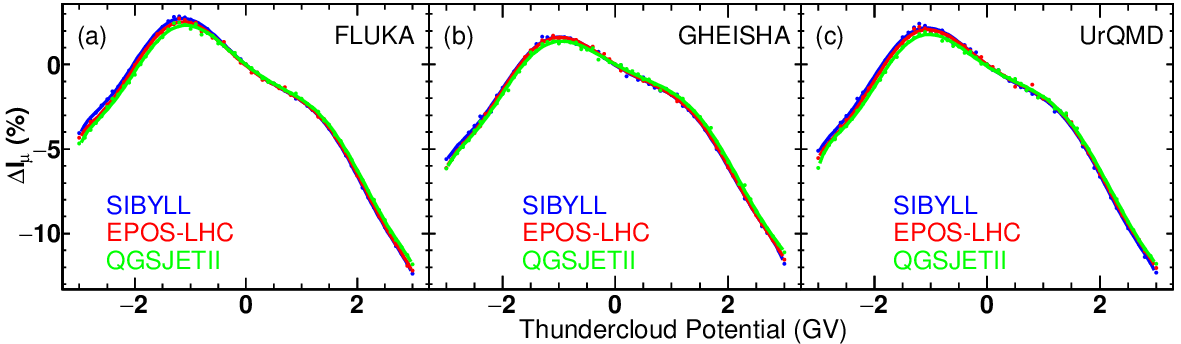}
    \caption{Simulated profiles of percent variation in muon intensity (\imu{}) among high-energy hadronic interaction generators of SIBYLL, EPOS-LHC, and QGSJETII are shown with reference to common low-energy hadronic interaction generator (a) FLUKA, (b) GHEISHA, and (c) UrQMD.}
    \label{Fig_5}
\end{figure*}

\begin{table}[b]
    \centering
    \begin{tabular}{|c|c|c|c|}
    \hline  
                & FLUKA     & GHEISHA   & UrQMD     \\
    \hline  
    SIBYLL      & 0.90      & 1.06      & 0.96      \\
                & ---       & (18\%)    & (7\%)     \\
    \hline  
    EPOS-LHC    & 0.93      & 1.08      & 1.00      \\
                & (3\%)     & (20\%)    & (11\%)    \\
    \hline  
    QGSJETII    & 0.95      & 1.13      & 1.00      \\
                & (5\%)     & (26\%)    & (11\%)    \\
    \hline  
    \end{tabular}
    \caption{Thundercloud potential \textbf{V} (GV) required to cause \imu=--2\% change in muon intensity for thunderstorm event of 1 December 2014 for nine different combinations of generators. Columns and rows represent low- and high-energy hadronic interaction generators, respectively, and values within parentheses show percent difference in thundercloud potential relative to SIBYLL-FLUKA combination.}
    \label{Tab_1}
\end{table}

\begin{figure*}[t!]
    \centering
    \includegraphics*[width=0.95\textwidth]{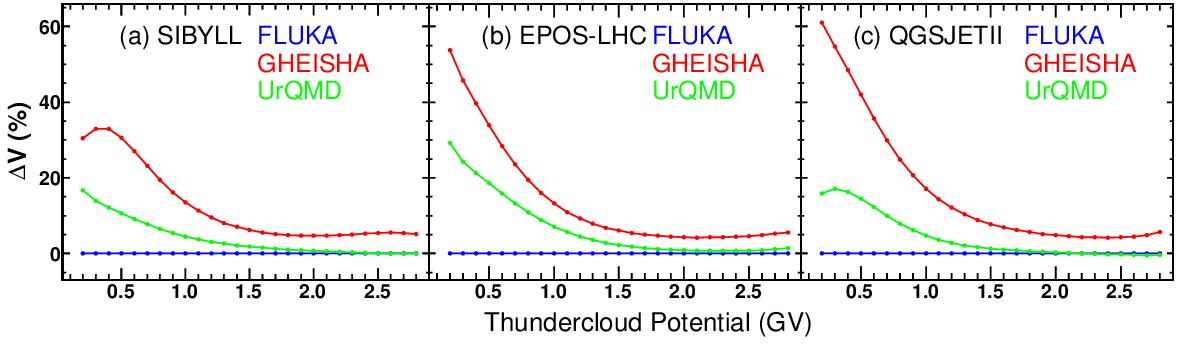}
    \caption{Simulated profiles of low-energy hadronic interaction generators FLUKA, GHEISHA, and UrQMD are shown with common high-energy hadronic interaction generators (a) SIBYLL, (b) EPOS-LHC, and (c) QGSJETII. Here, the ordinate in each sub-figure shows the difference in thundercloud potential relative to FLUKA.}
    \label{Fig_6}
    \vspace{1cm}
    \includegraphics*[width=0.95\textwidth]{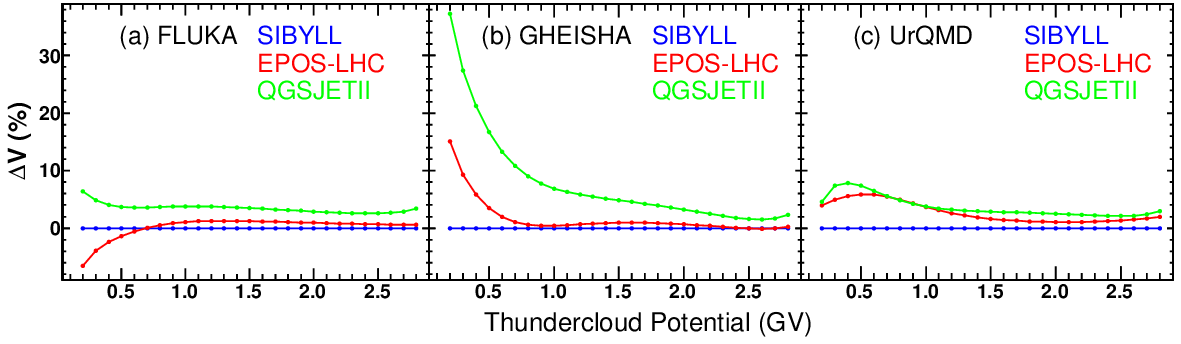}
    \caption{Simulated profiles of high-energy hadronic interaction generators SIBYLL, EPOS-LHC, and QGSJETII are shown with common low-energy hadronic interaction generators (a) FLUKA, (b) GHEISHA, and (c) UrQMD. Here, the ordinate in each sub-figure shows the difference in thundercloud potential relative to SIBYLL.}
    \label{Fig_7}
\end{figure*}

In Figure\,\ref{Fig_4} the \imu{} is shown as a function of applied potential for the low-energy generators, with one of the high-energy generators as a reference. Similarly, Figure\,\ref{Fig_5} shows the variation for high-energy generators after selecting one of the low-energy generators as the reference. It can be seen clearly that the low-energy generators have significantly large differences, especially for \textbf{--V}. However, the differences at \textbf{+V} are smaller compared to \textbf{--V}. Interestingly, the variations in the high-energy generators are not as significant when compared to the low-energy generators. This is most likely due to the transition energy of 80\,GeV from the high- to low-energy generators defined in CORSIKA. The bulk of CRs ($\sim$97\%) in the energy range of 10\,GeV--10\,TeV used here fall below 80\,GeV due to the power-law spectrum of CRs. Due to this fact, most of the hadronic interactions occur at energies below 80\,GeV resulting in dominant use of low-energy generators. From Figures\,\ref{Fig_4} \& \ref{Fig_5} the profiles of nine combinations were further analyzed quantitatively by using the observed muon intensity. Table\,\ref{Tab_1} lists the thundercloud potential required to produce observed peak muon deficit of 2\% for the event of 1 December 2014 (Figure\,\ref{Fig_3}). Here, the SIBYLL-FLUKA combination (used in earlier work) was used as a reference. The percent differences in thundercloud potential relative to the SIBYLL-FLUKA combination are displayed in the cells for the remaining combinations. Clearly, the SIBYLL-FLUKA combination yielded the lowest potential. The maximum deviation is found with the QGSJETII-GHEISHA (26\% higher) combination. On the other hand, the deviations are only a few percent ($\leqslant$8\%) among high-energy generators, irrespective of the low-energy generators used. However, the variations among low-energy generators are found to be as large as 21\%, especially when paired with QGSJETII. Overall, GHEISHA provides a higher estimate, whereas FLUKA yields the lowest estimate of potential among the low-energy generators. In summary, the SIBYLL-FLUKA combination yields the lowest estimate of thundercloud potential and is therefore, the most conservative estimate of thundercloud potential \cite{Hari_2019_2}. 

\begin{sidewaystable}
    \centering
    \begin{tabular}{|c|c|c|c|c|c|c|c|c|c|c|c|c|}
    \hline  
    \multicolumn{4}{|c|}{Thunderstorm event details} & \multicolumn{9}{|c|}{Thundercloud potential (GV)} \\
    \multicolumn{4}{|c|}{} & \multicolumn{9}{|c|}{(Percent change with respect to SIBYLL-FLUKA)} \\
    \hline
    \multirow{2}{*}{Date} & \multirow{2}{1.6cm}{No. of directions} & \multirow{2}{1.2cm}{\imu (\%)} & \multirow{2}{1.4cm}{Duration (min.)} & \multicolumn{3}{|c|}{SIBYLL} & \multicolumn{3}{|c|}{EPOS-LHC} & \multicolumn{3}{|c|}{QGSJETII} \\
    \cline{5-13}
    & & & & FLUKA & GHEISHA & UrQMD & FLUKA & GHEISHA & UrQMD & FLUKA & GHEISHA & UrQMD \\
    \hline
    13-10-2012       & 28 & --1.77 & 52 &  0.92   & 1.09   & 0.98   & 1.01   & 1.13   & 1.00   & 1.01   & 1.11   & 1.04    \\
                     &    &        &    &   ---   & (18\%) & (6\%)  & (9\%)  & (22\%) & (8\%)  & (10\%) & (20\%) & (13\%)  \\
    \hline
    11-04-2014       & 27 & --1.71 & 43 &  0.76   & 0.95   & 0.83   & 0.90   & 0.93   & 0.90   & 0.89   & 0.95   & 0.88    \\
                     &    &        &    &  ---    & (25\%) & (9\%)  & (18\%) & (22\%) & (18\%) & (17\%) & (25\%) & (16\%)  \\
    \hline
    23-09-2014       & 20 & --2.91 & 28 &  1.17   & 1.29   & 1.23   & 1.19   & 1.33   & 1.23   & 1.26   & 1.36   & 1.28    \\
                     &    &        &    &  ---    & (11\%) & (5\%)  & (1\%)  & (14\%) & (6\%)  & (8\%)  & (16\%) & (9\%)   \\
    \hline
    28-09-2014       & 39 & --2.03 & 19 &  1.12	  & 1.21   & 1.18   & 1.17   & 1.22   & 1.17   & 1.20   & 1.19   & 1.20    \\
                     &    &        &    &  ---    & (8\%)  & (5\%)  & (5\%)  & (9\%)  & (4\%)  & (7\%)  & (6\%)  & (7\%)   \\
    \hline
    01-12-2014       & 45 & --1.95 & 18 &  0.90   & 1.06   & 0.96   & 0.93   & 1.08   & 1.00   & 0.95    & 1.13   & 1.00   \\
                     &    &        &    &  ---    & (18\%) & (7\%)  & (3\%)  & (20\%) & (11\%) & (5\%)   & (26\%) & (11\%) \\
    \hline
    06-04-2017       & 12 & --3.04 & 22 &  1.39   & 1.46   & 1.42   & 1.35   & 1.44   & 1.42   & 1.45    & 1.55   & 1.47   \\
                     &    &        &    &  ---    & (5\%)  & (2\%)  & (--3\%) & (4\%) & (2\%)  & (4\%)   & (11\%) & (6\%)  \\
    \hline
    06-02-2018       & 40 & --2.00 & 47 &  1.04   & 1.13   & 1.07   & 1.13   & 1.15   & 1.11   & 1.04    & 1.18   & 1.11   \\
                     &    &        &    &  ---    & (8\%)  & (2\%)  & (8\%)  & (10\%) & (6\%)  & (0\%)   & (13\%) & (6\%)  \\
    \hline
    22-04-2018       & 29 & --1.64 & 35 &  0.86   & 0.99   & 0.88   & 0.94   & 0.97   & 0.95   & 0.97    & 1.03   & 0.95   \\
                     &    &        &    &  ---    & (16\%) & (3\%)  & (9\%)  & (13\%) & (12\%) & (13\%)  & (21\%) & (11\%) \\
    \hline  
    \end{tabular}
    \caption{List of major thunderstorm events recorded from April 2011 to December 2020, and the corresponding number of affected directions, peak change in muon intensity (\imu) (\%), event duration (min.), and thundercloud potentials (GV) derived from all the nine combinations of hadronic interaction generators. Values within parentheses show percent difference of thundercloud potential ($\Delta$V(\%)) relative to SIBYLL-FLUKA combination.}
    \label{Tab_2}
\end{sidewaystable}

\begin{table}[t]
    \centering
    \begin{tabular}{|c|c|c|c|}
    \hline  
                & FLUKA         & GHEISHA       & UrQMD         \\
    \hline  
    SIBYLL      & ---           & (14$\pm$7)\%  & (5$\pm$2)\%   \\
    \hline  
    EPOS-LHC    & (6$\pm$6)\%   & (14$\pm$7)\%  & (8$\pm$5)\%   \\
    \hline  
    QGSJETII    & (8$\pm$5)\%   & (17$\pm$7)\%  & (10$\pm$4)\%  \\
    \hline  
    \end{tabular}
    \caption{Summary of variations among the nine hadronic interaction generators derived from eight major thunderstorm events.}
    \label{Tab_3}
\end{table}

Although the differences across the hadronic interaction generators are significantly higher as seen in Table\,\ref{Tab_1}, the differences in Figures\,\ref{Fig_4} \& \ref{Fig_5} are barely noticeable due to the stronger dependence of \imu over a broader range of applied thundercloud potentials. Thus, these figures are redrawn with different perspectives to highlight the differences among different combinations. Here, we adopted a similar approach to that used for the presentation in Table\,\ref{Tab_1}. Figure\,\ref{Fig_6} shows the variation of thundercloud potential among low-energy generators, each paired with a high-energy generator. Since the SIBYLL-FLUKA combination is known to provide the most conservative estimates, FLUKA is kept as the common low-energy generator, with a different high-energy generator used for each sub-figure shown in Figure\,\ref{Fig_6}. For each step of applied thundercloud potential, the relative variation ($\Delta V$(\%)) relative to the reference combination is shown on the ordinate. The low-energy generators exhibit large variations among themselves. It is also noteworthy that the variations are significantly larger for smaller potentials. Unlike Table\,\ref{Tab_1} where the differences are compared for a single thundercloud potential, here the variations can be observed in a larger span of the thundercloud potentials. Similarly, Figure\,\ref{Fig_7} shows the variations in thundercloud potential caused by the choice of high-energy generators by keeping SIBYLL as the reference. Evidently, high-energy generators do not exhibit sizable variations, except in the cases when paired with GHEISHA. 

From the extended database of 487 significant thunderstorm events from April 2011 to December 2020 \cite{Hariharan_2021}, a list of eight major thunderstorm events, including the 1 December 2024, was prepared based on the maximum change of muon intensity and a relatively simple profile during the event. These events were analyzed following the same procedure as discussed in Section\,\ref{Event}, except for the part that involved tracking of cloud movement, which would have required a less turbulent electric field for successful analysis. Table\,\ref{Tab_2} lists the summary of these eight events with the total number of affected directions, the maximum change in muon intensity, the event duration, and the estimated thundercloud potentials derived for all nine combinations of hadronic interaction generators through Monte Carlo simulations in the subsequent columns. For each event, the relative difference in the thundercloud potential with respect to the SIBYLL-FLUKA combination is quoted in parentheses. To obtain a quantitative estimate of this dependence, a summary of the relative variations in thundercloud potential between these events for each combination is shown in Table\,\ref{Tab_3}. It is clear that the low-energy hadronic interaction generators display larger variations ($\sim$14\%), when compared to the high-energy generators ($\sim$8\%). Also, it should be noted that QGSJETII and GHEISHA provide systematically higher thundercloud potentials. Similarly, SIBYLL-FLUKA provide systematically lower thundercloud potentials.

\textbf{The conclusion that the SIBYLL-FLUKA combination provides the lowest estimates of thundercloud potentials was initially derived from a comprehensive simulation dataset generated using CORSIKA v74001 with all nine available hadronic event generator combinations. However, as hadronic interaction models continue to be updated and refined, it is essential to assess the impact of these changes on the estimated thundercloud potentials. To evaluate the variability of the SIBYLL-FLUKA combination, we conducted additional simulations using the latest CORSIKA release, v78000, interfaced with SIBYLL 2.3e and FLUKA 2024.1. These simulations covered a thundercloud potential range of --3 GV to +3 GV, employing a coarser step size (0.25\,GV instead of 0.1\,GV in previous simulations) and reduced statistics ($\sim$10\% of the original dataset for each direction and thundercloud potential step). A comparative analysis of eight thunderstorm events revealed that the estimated thundercloud potentials differed by --(0.3$\pm$4.8)\% between the two CORSIKA versions. The small mean variation (--0.3\%) and standard deviation (4.8\%) suggest that these differences lie within the measurement uncertainties (i.e., the 10\% error on the estimated thundercloud potential, as shown in Figure\,5 of \cite{Hari_2019_1}) and are statistically insignificant. Therefore, the thundercloud potentials derived using the SIBYLL-FLUKA combination, which consistently yielded the lowest values for the eight events analyzed in this study, remain robust and well-justified.}


\section{Discussion \label{Discussion}}

More than two centuries of research in thunderstorm physics resulted in numerous discoveries and continually improving understanding of various aspects of thunderstorms. Much of the advancement on the experimental front occurred in the past few decades primarily because the technological advances in measurement of thunderstorm properties. Nevertheless, we still lack tools powerful enough to fully understand thunderstorm physics. In the present work, the close connection between CRs and thunderstorms has been exploited to improve our understanding of thunderstorm properties. The charged secondaries produced by CRs in the atmosphere serve as key messengers, that get affected by the thundercloud potential. In particular, muons are the ideal choice for such studies because of their constant energy loss rate in the atmosphere and their charge asymmetry. The muon charge asymmetry is an outcome of the combined effects of the geomagnetic field on primary and secondary CRs and predominantly positive charge of CRs. Due to this asymmetry, the thundercloud potential causes a net change in the muon intensity. A ground-based experiment with accurate measurement of the muon intensity can provide an estimate of thundercloud potential.

The G3MT is a ground-based instrument that is well suited for the study of thunderstorms. It is capable of measuring the variation in angular muon intensity (E$_\mu\geqslant$1\,GeV) with a precision of $\sim$0.1\% on timescale of minutes. The G3MT had estimated an electric potential of 1.3\,GV in one of the biggest thunderstorm events recorded on 1 December 2014, which confirmed an almost a century-old prediction by C.T.R. Wilson, superseding the earlier measurements by an order of magnitude carried out by sounding techniques. This estimation also offers an understanding the origin of high-energy photons detected in the terrestrial gamma-ray flashes by space probes. As discussed before, the estimations were based on Monte Carlo simulations, which are inherently dependent on hadronic interaction generators used in the development of EAS. It is shown that these generators produce minor differences in the magnitude of estimated thundercloud potential, even after tuning them with LHC data \cite{Pierog_2017}. As mentioned before, these generators are built on assumptions based on a simple parton models associated with the Gribov-Regge multiple scattering. However, the implementation of physics treatment among these generators via different approaches leads to variation in the selection effects of pion, kaon, and muon production. As a result, notable differences are found among the predictions of these generators. Significant differences in the estimated thundercloud potential are observed among the nine combinations of hadronic interaction generators, derived from three generators each for the high- and low-energy segments, respectively. In the present work, eight major thunderstorm events observed by G3MT from April 2011 to December 2020 have been analyzed and their findings are reported. The high-energy generators cause a variation of $\sim$8\%, whereas the low-energy generators produce variation of $\sim$14\%. As the interactions below 80\,GeV are handled by low-energy generators and a large majority of CRs are below this energy in the selected energy range of 10\,GeV--10\,TeV, the difference in low-energy generators dominate. The largest difference of 17\% occurs for QGSJETII-GHEISHA with respect to the SIBYLL-FLUKA combination. \textbf{Given ongoing updates to hadronic event generators, we reassessed this conclusion using CORSIKA v78000 with SIBYLL 2.3e and FLUKA 2024.1. A comparison of eight thunderstorm events showed a deviation of --(0.3$\pm$4.8)\%, which falls within the 10\% measurement uncertainty (Figure 5 of \cite{Hari_2019_1}). These findings confirm that the SIBYLL-FLUKA combination remains the most consistent and  conservative estimator of thundercloud potentials.} Thus, the previously reported thundercloud potential of 1.3\,GV is the most conservative estimate for the event of 1 December 2014 by using SIBYLL-FLUKA. Furthermore, this choice of SIBYLL-FLUKA generators also provides the lowest estimate of thundercloud potential for the other seven major thunderstorm events. This work provides a framework for estimating the thundercloud properties with the uncertainties introduced by different hadronic interaction generators used in CORSIKA, which are especially important for future such studies.


\section{Summary \label{Summary}}

The study by the GRAPES-3 muon telescope (G3MT) demonstrates that,

\begin{enumerate}
    \item The estimated thundercloud potential depends on the choice of hadronic interaction generators used in the Monte Carlo simulations of atmospheric muons.
    \item Among the nine combinations of (three low-energy and three high-energy) generators, the SIBYLL-FLUKA combination invariably yielded the lowest thundercloud potential, which is reported in order to be conservative.
    \item Low-energy generators produce significantly larger variations in thundercloud potential than the high-energy generators due to the dominance of interactions at low-energies.
    \item The G3MT is a continuous sky monitoring system, recording $\sim$50 significant thunderstorms annually. The analysis of a decade-long dataset, revealed eight major thunderstorms of gigavolt potential, underscoring the generation of such potentials as their characteristic feature.
\end{enumerate}


\section{Future directions \label{Future}}

\textbf{The G3MT continues to record approximately 50 thunderstorm events annually. A comprehensive analysis of the entire dataset collected since its inception is essential for establishing the statistical distribution of thundercloud properties. This long-term study can offer insights into seasonal and interannual variations in thundercloud characteristics, contributing to a broader understanding of atmospheric electricity. Monte Carlo simulations remain indispensable for probing these properties using G3MT. This study has demonstrated the influence of hadronic event generators on the estimation of thundercloud properties. Future work may also explore the role of thundercloud geometry in simulations to refine these estimations further. Additionally, G3MT has been collecting high-resolution electric field measurements using four widely spaced EFMs over the past decade. This extensive dataset, combined with muon observations, presents a valuable opportunity for interdisciplinary research. Leveraging modern computational techniques, such as machine learning, to analyze the low-level data could provide deeper insights into the underlying physical processes, and develop predictive models for thunderstorm-related atmospheric phenomena.}
\\


\section*{Acknowledgement}

We thank D.B. Arjunan, V. Jeyakumar, S. Kingston, K. Manjunath, S. Murugapandian, S. Pandurangan, B. Rajesh, V. Santhoshkumar, M.S. Shareef, C. Shobana, and R. Sureshkumar for their efforts in maintaining the GRAPES-3 experiment. The authors also express sincere thanks to the remaining members of the collaboration for their review and comments. We acknowledge the support of the Department of Atomic Energy, Government of India, under Project Identification No. RTI4002. \textbf{We sincerely appreciate the referee's thorough review and thoughtful suggestions, which have greatly contributed to the improvement of this manuscript. Their insightful feedback has helped refine our analysis and enhance the clarity of our presentation.}


\end{document}